\documentclass[prc,showpacs,showkeys,superscriptaddress,nofootinbib,twocolumn,floatfix]{revtex4-1}
\usepackage{amsfonts}
\usepackage{float}
\usepackage{placeins}
\usepackage{graphicx}
\usepackage[hidelinks]{hyperref}
\usepackage{color,amsmath,amssymb,bm}

\def\blfootnote{\xdef\@thefnmark{}\@footnotetext}
\def\eg{{\em e.g.}}
\def\ie{{\em i.e.}}

\newcommand{\beq}{\begin{equation}}
\newcommand{\eeq}{\end{equation}}
\newcommand{\bea}{\begin{eqnarray}}
\newcommand{\eea}{\end{eqnarray}}

\newcommand{\gtsim}{\raisebox{-4pt}{$\,\stackrel{\textstyle >}{\sim}\,$}}

\begin{document}

\title{Probing the in-Medium QCD Force by Open Heavy-Flavor Observables}
\author{Shuai Y.F.~Liu}
\affiliation{Cyclotron Institute and Department of
Physics and Astronomy, Texas A\&M University, College Station, TX 77843-3366, USA}
\author{Min He}
\affiliation{Department of Applied Physics, Nanjing University of Science and Technology, 
Nanjing 210094, China} 
\author{Ralf Rapp}
\affiliation{Cyclotron Institute and Department of
Physics and Astronomy, Texas A\&M University, College Station, TX 77843-3366, USA}

\date{\today}

\begin{abstract}
The determination of the color force in a quark-gluon plasma (QGP) is a key objective in the 
investigation of strong-interaction matter. Open and hidden heavy-flavor observables in heavy-ion 
collisions (HICs) are believed to provide insights into this problem by comparing calculations 
of heavy-quark (HQ) and quarkonium transport with pertinent experimental data. In this work, we 
utilize the $T$-matrix formalism to compute charm-quark transport coefficients for various input 
potentials previously extracted from simultaneous fits to lattice-QCD data for HQ free energies, 
quarkonium correlators and the QGP equation of state. We investigate the impact of off-shell 
effects (spectral functions) in the QGP medium on the HQ transport, and compare to earlier 
results using the free or internal HQ energies as potential proxies. We then employ the 
transport coefficients in relativistic Langevin simulations for HICs to test the sensitivity of 
heavy-flavor observables to the HQ interactions in the QGP. We find that a strongly-coupled 
$T$-matrix solution generates a HQ elliptic flow comparable to the results from the internal 
energy at low momentum, driven by a long-range remnant of the confining force, while falling 
off stronger with increasing 3-momentum. The weakly coupled $T$-matrix solution, whose underlying 
potential is close to the free energy, leads to an elliptic flow well below the experimentally 
observed range.  
\end{abstract}

\pacs{25.75.Dw, 12.38.Mh, 25.75.Nq}
\keywords{Heavy-Flavor Transport, Quark Gluon Plasma, Ultrarelativistic Heavy-Ion Collisions}

\maketitle

%%%%%%%%%%%%%%%%%%%%%%%%%%%%%%%%%%%%%%%%%%%%%%%%%%%%%%%%
\section{Introduction}
\label{sec_intro}
%%%%%%%%%%%%%%%%%%%%%%%%%%%%%%%%%%%%%%%%%%%%%%%%%%%%%%%
The investigation of the in-medium color force between partons is pivotal for understanding 
the microscopic mechanisms that lead to the remarkable features of the quark-gluon plasma (QGP) 
as observed in ultra-relativistic heavy-ion collisions (URHICs). 
Lattice-QCD (lQCD) computations of the free energy of a heavy quark-antiquark ($Q\bar Q$) pair 
immersed into the QGP~\cite{Petreczky:2004pz,kaczmarek2005static}
%\cite{Datta:2003ww,Petreczky:2004pz,kaczmarek2005static,Aarts:2007pk,Aarts:2011sm}
indicate that nonperturbative effects, specifically remnants of the linear part of the potential, 
survive up to temperatures of at least twice the pseudo-critical one, 
$T_{\rm pc}\simeq 160$\,MeV~\cite{Borsanyi:2010bp,Bazavov:2011nk}.
%\cite{Aoki:2006br,Aoki:2009sc,Borsanyi:2010bp,Bazavov:2011nk} 
Potential models~\cite{Wong:2004zr,Cabrera:2006wh,Alberico:2006vw,Mocsy:2007yj,Riek:2010py} have 
been employed to implement these effects and test them against lQCD data 
for euclidean quarkonium correlators~\cite{Jakovac:2006sf,Aarts:2007pk,Ding:2012sp,Aarts:2014cda},
%,Bazavov:2014cta
but no definite answer on the modifications of the QCD force in medium could be achieved.
To broaden these investigations we have been developing a thermodynamic $T$-matrix 
approach~\cite{Mannarelli:2005pz,Riek:2010py,Liu:2016ysz,Liu:2017qah} where 
consequences of the in-medium potential are assessed not only for quarkonia, but also for 
individual heavy quarks (such
as their transport properties) and the surrounding medium that they interact with.  
The $T$-matrix framework has been solved selfconsistently for one- and two-parton 
correlations in a full off-shell scheme beyond the quasiparticle 
approximation~\cite{Liu:2016ysz,Liu:2017qah}, allowing 
for the dynamical formation of (broad) bound states, and connecting bulk and microscopic 
properties of the QGP and its excitations (spectral functions). Despite constraints from 
three sets of lQCD data (equation of state (EoS), heavy-quark (HQ) free energy, and 
quarkonium correlators), the underlying in-medium potential could still not be determined
unambiguously~\cite{Liu:2017qah}. However, different potentials predict markedly 
different spectral and transport properties of the QGP. 
The objective of the present paper is to further explore this sensitivity by computing the 
thermal relaxation rates for charm quarks for different potential solutions (including 
previously used internal- and free-energy proxies) and quantifying their effect on the 
charm-quark spectra in URHICs using relativistic Langevin simulations. We specifically
scrutinize off-shell effects in the calculation of the transport coefficients, which can 
play a significant role given the large spectral widths of partons found in the 
``strongly-coupled solution" of the $T$-matrix approach, together with broad $D$-meson 
resonance states in the charm-light-quark scattering amplitude near or even below the 
nominal two-parton threshold.

This paper is organized as follows. In Sec.~\ref{sec_pot} we recollect the main features 
and differences of weakly- and strongly-coupled solutions that we previously found within 
the $T$-matrix approach. In Sec.~\ref{sec_offtrans} we introduce the off-shell formalism 
to calculate HQ transport coefficients, and discuss an improved partial-wave expansion in 
the $T$-matrix over previous calculations of the HQ relaxation rate.
In Sec.~\ref{sec_trans} we analyze the results of the HQ transport coefficients
from the different types of potentials. In Sec.~\ref{sec_urhic}, we briefly recall
the transport implementation into URHICs using relativistic Langevin 
simulations, calculate the charm-quark and $D$-meson nuclear modification factors ($R_{AA}$) 
and elliptic flow ($v_2$), and discuss the results in light of discriminating different 
potential strengths via experimental observables.
In Sec.~\ref{sec_concl} we summarize and conclude.
In appendix~\ref{app_cm}, we collect the expressions used for the transformation of the 
off-shell $T$-matrix into the center-of-mass (CM) frame as used in this work.

%%%%%%%%%%%%%%%%%%%%%%%%%%%%%%%%%%%%%%%%%%%%%%%%%%%%%%%%%%%%%%%%
%\section{Heavy-Flavor Transport Coefficients}
%\label{sec_trans}
%%%%%%%%%%%%%%%%%%%%%%%%%%%%%%%%%%%%%%%%%%%%%%%%%%%%%%%%%%%%%%%%
\section{In-Medium Potentials Based on Lattice QCD}
\label{sec_pot}
\begin{figure} [thb!]
	\centering
	\includegraphics[width=0.99\columnwidth]{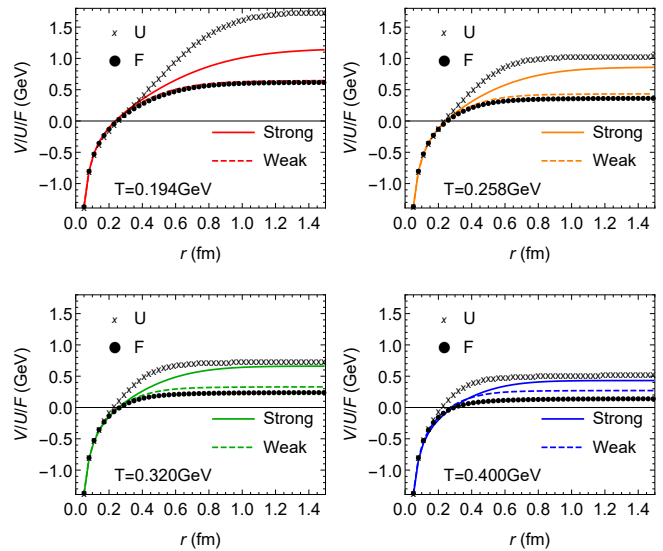}
\caption{The potentials of the SCS (solid lines) and WCS (dashed lines) are compared to the 
internal energy $U$ (crosses) and free energy $F$ (dots) as a function of distance between
a $Q$ and $\bar Q$, for four temperatures.
	}
	\label{fig_pot}
\end{figure}

%%%%%%%%%%%%%%%%%%%%%%%%%%%%%%%%%%%%%%%%%%%%%%%%%%%%%%%%%%%%%%%%
In Ref.~\cite{Liu:2017qah} we deployed the $T$-matrix approach, together with the
Luttinger-Ward Baym formalism, in a comprehensive fit
to lQCD data for the HQ free energy, quarkonium correlators and the QGP EoS.
It turned out that the input potential required to simultaneously describe the lattice data 
is not unique. We bracketed the range of viable potentials by approximately limiting 
scenarios referred to as a strong-coupling scenario (SCS) and a weak-coupling
scenario (WCS). The main features of the SCS are large thermal parton widths leading 
to a dissolution of their quasiparticle peaks at low momentum near $T_{\rm pc}$ while 
broad mesonic and diquark bound states emerge whose contribution dominates the pressure 
when approaching $T_{\rm pc}$ from above.
On the other hand, in the WCS thermal partons remain well defined quasiparticles (with
masses similar to the SCS), while rather narrow, loosely-bound two-body states form near 
$T_{\rm pc}$ whose contribution to the 
EoS remains, however,  subleading. The underlying static potentials, $V_s$ for the SCS
and $V_w$ for the WCS, are displayed in Fig.~\ref{fig_pot}, along with the lQCD results 
for the free ($F$) and internal energy ($U$) that they reproduce through the $T$-matrix 
formalism. Both ${V}_s$ and ${V}_w$ lie in between $U$ and $F$, and they tend to be closer 
to $U$ as temperature increases while their gap diminishes. However, at low temperatures 
$V_w$ essentially coincides
with $F$ while $V_s$ reaches much above it at intermediate and especially large distances. 
This difference is the key factor in the resulting QGP spectral properties near $T_{\rm pc}$ 
as discussed above; the large-distance strength of $V_s$ implies that the QGP is strongly 
coupled only at large distances, \ie, for soft momenta. 
%A more detailed comparisons for the $V$, $U$ and $F$ and their impacts on transport 
%coefficients will be discussed in Sec*.~\ref{sec_vufcompare}.

\begin{figure} [!htb]
	\centering
	\includegraphics[width=0.99\columnwidth]{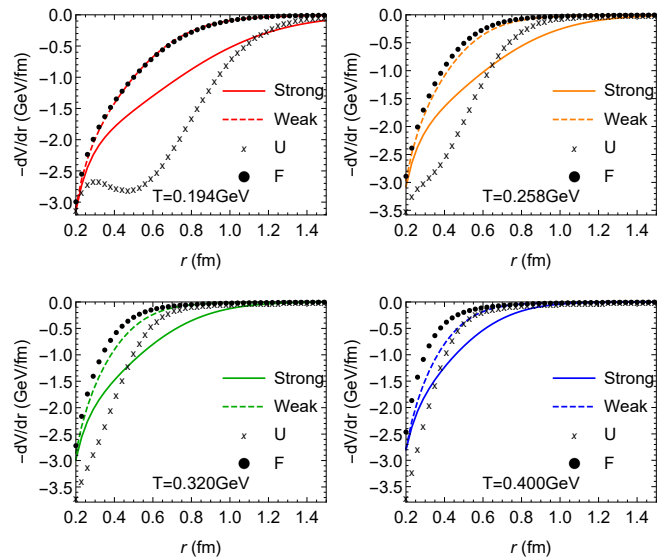}
	\caption{Force for $V_s$ (solid line), $V_w$ (dashed line), $U$ (crosses) and $F$ (dots) at different temperatures.}
	\label{fig_vufforce}
\end{figure}
\begin{figure} [!htb]
	\centering
	\includegraphics[width=0.99\columnwidth]{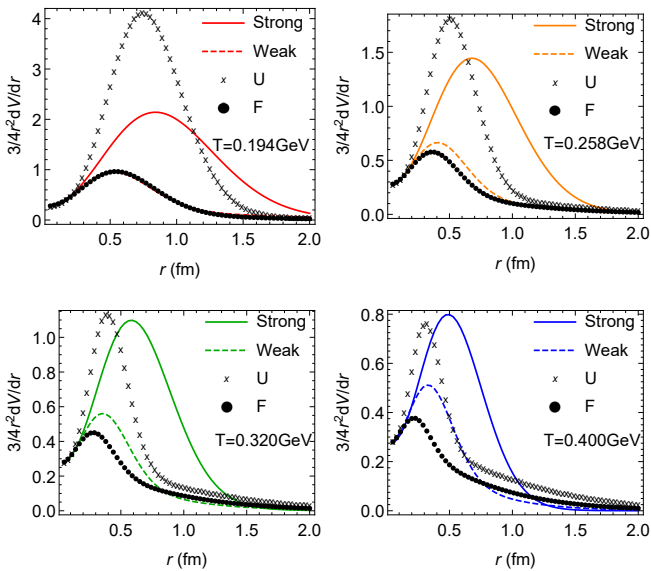}
	\caption{The dimensionless quantity $\frac{3}{4}r^2 dV/dr$ (scaled to recover the strong 
coupling constant, $\alpha_s$, at short distance) is plotted for $V_s$ (solid line), $V_w$ 
(dashed line), $U$ (crosses) and $F$ (dots).}
	\label{fig_vasr}
\end{figure}

\begin{figure} [!htb]
	\centering
	\includegraphics[width=0.99\columnwidth]{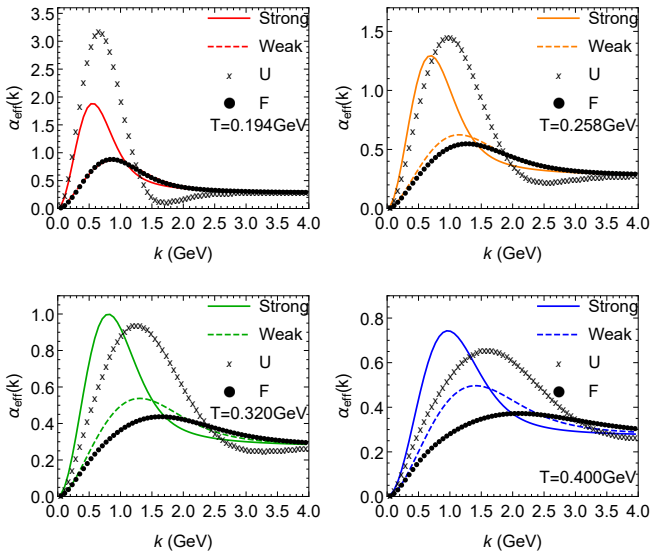}
	\caption{The dimensionless quantity $\alpha_{\rm eff}(k)\equiv\frac{3}{16\pi}k^2 V(k)$ (scaled 
to recover the strong coupling constant, $\alpha_s$, at large momentum) is plotted for  $V_s$ (solid 
line), $V_w$ (dashed line), $U$ (crosses) and $F$ (dots).}
	\label{fig_vasp}
\end{figure}

Taking the derivative of the potentials, $-dV(r)/dr$, yields the pertinent forces,
cf.~Fig.~\ref{fig_vufforce}. The forces for $V_s$ and $U$ at large distances are much higher 
than those for $ V_w $ and $F$. Around $r\simeq0.5$\,fm and $T$=0.194~GeV, the force from $U$ 
amounts to ca.~2.5\,GeV/fm which even exceeds the vacuum force by about a factor of $\sim$2. 
This enhancement originates from the ``entropy term", $-TdF/dT$, as a fast change in 
degrees of freedom near $T_{\rm pc}$ leads to a large temperature derivative. It has been
suggested that this is caused by releasing thermal magnetic monopoles~\cite{Liao:2008vj}. 
The force from $V_s$ at this distance (at $T$=0.194~GeV) is also larger than in vacuum,
by about 20\%; \ie, the major contribution to this force is still considered 
to be the remnant of the confining vacuum configuration rather than thermal monopoles.

The long-range force is closely related with low-momentum transport properties of
the medium; in particular, a long-range force allows a parton to interact with an increased 
number of thermal partons in the heat bath, proportional to the volume of the spherical shell 
which grows as $r^2$. Therefore, by multiplying the force with $\frac{3}{4}r^2$, one forms a 
dimensionless quantity, $\frac{3}{4} r^2 dV/dr$, that can be regarded as an ``effective 
interaction strength" in the medium
%~\footnote{Due to different origins of the Coulomb and confining terms (different relativistic 
%and color structure), we should be careful about the concept of ``effective coupling".} 
and is plotted for the 4 ``potentials" in Fig.~\ref{fig_vasr}. The factor of 3/4 renders 
the $r\to0$ limit equal to the strong coupling constant, which is $\alpha_s$=0.27 for all
of our 4 ``potentials". Starting from short range, $U$  has the largest interactions, 
up to $r\simeq 1(0.4)$\,fm at the smallest (largest) temperature, due to the ``entropy-related" 
potential, $-TdF/dT$; as we will see below, this can affect transport properties
even at rather high momentum.
Coming from the large distance side, $V_s$ gives the strongest ``effective" coupling, and
its maximum coupling peak at each temperature is located at the largest distance among all 
potentials, ranging from $r_{\rm max}$=0.85\,fm at $T$=0.194\,GeV down to $r_{\rm max}$=0.5\,fm 
at $T$=0.400\,GeV. The large-distance enhancement of the coupling can be related to an infrared 
enhancement in momentum space, as illustrated by the dimensionless-scaled momentum space potentials 
displayed in Fig.~\ref{fig_vasp}: here, the maximum interaction strength for $V_s$ occurs at the 
lowest momentum (relative momentum exchange between $Q$ and $\bar Q$) among the 4 potentials, 
approximately given by $p_{\rm max}=2/r_{\rm max}$.

%%%%%%%%%%%%%%%%%%%%%%%%%%%%%%%%%%%%%%%%%%%%%%%%%%%%%%%%%%%%
\section{Off-Shell Transport Coefficients}
\label{sec_offtrans}
%%%%%%%%%%%%%%%%%%%%%%%%%%%%%%%%%%%%%%%%%%%%%%%%%%%%%%%%%%%%
As mentioned above, the strong color force, in particular in the SCS, leads to large widths in 
the spectral functions of thermal partons, dissolving their quasiparticle peaks at 
low momenta and temperatures~\cite{Liu:2017qah}. It is therefore in order to incorporate 
the off-shell effects in the Boltzmann/Langevin description of the HQ transport. Toward this 
goal, we start from the Kadanoff-Baym equations and use a minimal set of approximations to 
reduce them to a Boltzmann equation, where quantum effects are encoded in the transition 
rates. Subsequently, this  Boltzmann equation is expanded into a Fokker-Planck equation, which 
can be implemented via a Langevin process where quantum effects are encoded in the transport 
coefficients.

We closely follow the formalism for non-equilibrium quantum field theory described in 
Ref.~\cite{Danielewicz:1982kk}. We first illustrate a formal derivation of the relations for 
the non-relativistic case, but our final expressions for the transport coefficients account 
for relativistic effects as discussed in Ref.~\cite{Liu:2017qah}. In relative energy-momentum 
space, with a macroscopic time denoted as \(t\),\footnote{We use the same approximation, 
\(T\pm t/2\approx T\), as in Ref.~\cite{Danielewicz:1982kk}, but use \(t\) to denote 
\(T=(t_1+t_2)/2\).} the equation for the non-equilibrium HQ Green function can be expressed 
as\footnote{We enforce translational invariance so that all terms with a coordinate gradient 
vanish, and the Boltzmann equation used to evaluate the transport coefficients 
can be obtained as in Ref.~\cite{Svetitsky:1987gq}.}
\begin{align}
&\frac{\partial}{\partial t}[\int d\omega G_Q^<(\omega,\textbf{p},t)]=
\int d\omega\{i\Sigma_Q^{<}(\omega,\textbf{p},t)G_Q^>(\omega,\textbf{p},t)
\nonumber\\
& \quad \qquad \qquad \qquad \qquad - i\Sigma_Q^{>}(\omega,\textbf{p},t)) G_Q^<(\omega,\textbf{p},t)\} \ .
\label{eq_baymeq} 
\end{align}
The $G_Q^{<,>}(\omega,\textbf{p},t)$ are the Fourier transforms of the Green functions,
\begin{align}
G_Q^{<}(t_1,x_1,t_2,x_2) &=i\langle\psi_Q^\dagger(t_2,x_2)\psi_Q(t_1,x_1)\rangle 
\\
%$G_Q^<(t_2,x_2,t_1,x_1)=\langle\ps 
G_Q^{>}(t_1,x_1,t_2,x_2) &=-i\langle\psi_Q(t_1,x_1)\psi^\dagger_Q(t_2,x_2)\rangle \ ,
%$G_Q^<(t_2,x_2,t_1,x_1)=\langle\psi_Q(t_1,x_1)\psi^\dagger_Q(t_2,x_2)\rangle$ 
\end{align}
with respect to $\delta t$ and $ \delta x $ for fixed $ t $ and $ x $ where 
$\delta t=t_1-t_2$, $\delta x=x_1-x_2$, $t=(t_1+t_2)/2$, $x=(x_1+x_2)/2$~\cite{Danielewicz:1982kk}.
In a uniform medium the $G_Q^{<,>}$ do not depend on $x$. 
$\Sigma_Q^{<,>}$ is the selfenergy in the real-time formalism, in which it can be calculated 
diagrammatically from the underlying scattering processes between the heavy quark and the 
partons of the medium. The Fourier transform of $\Sigma_Q^{<,>}$ uses the same convention as that 
for $G_Q^{<,>}$. The $T$-matrix approach has been used to derive the expressions for these 
selfenergies in Appendix F of Ref.~\cite{Danielewicz:1982kk}. One has 
\begin{align}
&\Sigma_Q^{>}(\omega,\textbf{p},t)=\mp \sum\int \frac{d\omega'd^4\textbf{p}'}{(2\pi)^4}\frac{d\nu d^3\textbf{q}}{(2\pi)^4}
\frac{d\nu'd^4\textbf{q}'}{(2\pi)^4}(2\pi)^4\delta^{(4)}  
\nonumber\\
& \quad \times |T(E,\textbf{P},\textbf{p},\textbf{p}')|^2 G^{>}_Q(\omega',p')G^{<}_i(\nu,q)G^{>}_i(\nu',q') 
 \ ,
\label{eq_selfT1}
\end{align}
and 
\begin{align}
&\Sigma_Q^{<}(\omega,\textbf{p},t)=\mp \sum\int \frac{d\omega'd^4\textbf{p}'}{(2\pi)^4}\frac{d\nu d^3\textbf{q}}{(2\pi)^4}
\frac{d\nu'd^4\textbf{q}'}{(2\pi)^4}(2\pi)^4\delta^{(4)}  
\nonumber\\
&\quad \times |T(E,\textbf{P},\textbf{p},\textbf{p}')|^2 G^{<}_Q(\omega',p')G^{>}_i(\nu,q)G^{<}_i(\nu',q')
\ .
%&|T(E,\textbf{P},\textbf{p}',\textbf{p})|^2 G^{<}_Q(\omega',p',t)G^{>}_i(\nu',q')G^{<}_i(\nu,q)
\label{eq_selfT2}
\end{align}
Here, $\delta^{(4)}$ is a short-hand notation for energy-momentum conservation, and $\sum$ 
represents the summation over the internal degrees of freedom $i=q, \bar{q}, g$ and their 
color, spin and flavor, divided 
by one HQ degeneracy, $d_Q$=6; $\textbf{P}$ and $E$ are the total momentum and energy, and
$T(E,\textbf{P},\textbf{p},\textbf{p}')$ is the retarded \(T\)-matrix. The $ G_{i}^{<,>} $ 
are the Green functions for the light partons in medium. The classical Boltzmann equation is 
recovered from Eq.~(\ref{eq_baymeq}) using the on-shell approximations: 
$G^<=\mp i(2\pi)\delta(\omega-\varepsilon(\textbf{p}))f(\textbf{p},t))$ and 
$G^>=-i (2\pi)\delta(\omega-\varepsilon(\textbf{p}))(1\pm f(\textbf{p},t))$.
These approximations are derived in Ref.~\cite{Danielewicz:1982kk};\footnote{Our 
convention for ``$ \mp $'' (upper/lower sign denotes boson/fermion) is opposite of that in 
Ref.~\cite{Danielewicz:1982kk}.} they neglect off-shell quantum effects, but not all
are necessary to describe HQ diffusion in a local-equilibrium QGP. We 
have found that ``minimal" approximations required for obtaining a HQ Boltzmann 
equation amount to
\begin{align}
&G_Q^< (\textbf{p},\omega,t)=i(2\pi)\delta(\omega-\varepsilon_Q(\textbf{p}))f_Q(\textbf{p},t) \ ,
\nonumber\\ 
&G_Q^{>}(\omega,p)=- i(2\pi)\rho_Q(\omega,p)(1- n_Q(\omega)) \ ,
\nonumber\\
&G_i^{<}(\omega,p)=\mp i(2\pi)\rho_i(\omega,p)n_i(\omega) \ ,
\nonumber\\
& G_i^{>}(\omega,p)=- i(2\pi)\rho_i(\omega,p)(1\pm n_i(\omega)) \ ,
\label{eq_approx}
\end{align}
\normalsize
where the quasiparticle approximation is only applied to $ G_Q^< (\omega,\textbf{p},t) $, \ie,
the incoming heavy quark, while all other $ G^{<,>} $ are taken to be off-shell equilibrium 
Green functions, with $\rho_{i,Q}$ and $ n_{i,Q} $ denoting the corresponding spectral and 
distribution functions, respectively, for light ($i$) and  heavy ($Q$) partons in equilibrium. 
Substituting these expressions into Eqs.~(\ref{eq_baymeq}), (\ref{eq_selfT1}), and (\ref{eq_selfT2}), 
yields the Boltzmann equation
\begin{align}
&\frac{\partial}{\partial t}f(\textbf{p},t)=
\nonumber\\
&\int \frac{d^3\textbf{k}}{(2\pi)^3} 
[w(\textbf{p+k},\textbf{k})f(\textbf{p+k},t)-w(\textbf{p},\textbf{k})f(\textbf{p},t)] \ ,
\end{align}
\normalsize
where the transition rate is\footnote{Note that 
$i\Sigma^{>}(p,\varepsilon(p),t)f(\textbf{p},t)=\int \frac{d^3\textbf{k}}{(2\pi)^3} 
[w(\textbf{p},\textbf{k})f(\textbf{p},t)]$. Also, when converting the gain term, 
$ \Sigma_Q^< G_Q^>$, to Boltzmann form, it is necessary to use 
$T(E,\textbf{P},\textbf{p},\textbf{p}')=T(E,\textbf{P},\textbf{p}',\textbf{p})$.}
\begin{align}
&w(\textbf{p},\textbf{k})=\int \frac{d\nu d^3\textbf{q}}{(2\pi)^3}\frac{d\nu'd^3\textbf{q}'}{(2\pi)^3}
d\omega'(2\pi)^4\delta^{(4)} \rho_i(\nu,q)
\nonumber\\
& \quad \qquad \times \rho_i(\nu',q') \rho_Q(\omega',|\textbf{k}+\textbf{p}|) 
|T(E,\textbf{P},\textbf{p},\textbf{k}+\textbf{p})|^2 
\nonumber\\
& \qquad \quad \qquad \qquad \times n_i(\nu) \left[1\mp n_i(\nu')\right] \left[1- n_Q(\omega')\right] \ ,
\label{eq_wrate}
\end{align}
\normalsize
and $\textbf{k}= \textbf{p}'-\textbf{p}$ is the 3-momentum exchange. Note that we have 
approximated the distribution function of the outgoing heavy quark in the blocking factor 
$(1-n_Q)$ to be a thermal one (the blocking factor is close to one in any case), and therefore 
the rate \(w(\textbf{p},\textbf{k})\) does not depend on the dynamical non-equilibrium 
HQ distribution function, \(f(\textbf{p},t)\).
So far, our discussion does not include relativistic effects; several modifications are necessary 
to do that, as detailed in the following for the calculation of the HQ transport coefficients.

Expanding the full Boltzmann equation in the momentum transfer, \(\textbf{k}\), results in 
a Fokker-Planck equation, which can be converted to a Langevin approach for heavy quarks. 
The Fokker-Planck equation is given by
\begin{align}
&\frac{\partial}{\partial t}f(p,t)=\frac{\partial}{\partial p_i}\{A_i(p)f(p,t)+
\frac{\partial}{\partial p_j}[B_{ij}(p)f(p,t)]\}
\end{align}
where the HQ transport coefficients are defined as weighted averages over the transition rate,
\begin{align}
&A_i(p)=\int \frac{d^3\textbf{k}}{(2\pi)^3} w(\textbf{p},\textbf{k}) k_i\nonumber\\
&B_{ij}(p)=\int \frac{1}{2}\frac{d^3\textbf{k}}{(2\pi)^3}w(\textbf{p},\textbf{k}) k_i k_j \ .
\end{align}
In local equilibrium, the drag ($A$) and transverse/longitudinal diffusion coefficients 
($B_0/B_1 $) are defined through
\begin{align}
&A_i(p)=A(p)p_i \nonumber\\
&B_{ij}(p)=B_0(p)P^{\perp}_{ij}+B_1(p)P^{\parallel}_{ij} \ ,
\end{align}
with the projectors $P^{\perp}_{ij}=\delta_{ij}-p_i p_j/\textbf{p}^2$ and 
$P^{\parallel}_{ij}=p_i p_j/\textbf{p}^2$. The scalar transport coefficients,
$A(p)$, $B_0(p)$ and $B_1(p)$, can thus be expressed via averages
\begin{equation}
\langle X(\mathbf{p}')\rangle\equiv\int \frac{d^3\mathbf{k}}{(2\pi)^3} w(\mathbf{p},\mathbf{k}) X(\mathbf{p}')
\end{equation}
as
\begin{eqnarray}
A(p)&=&\langle 1-\frac{\textbf{p}\cdot\textbf{p}'}{\textbf{p}^2}\rangle
\nonumber\\
B_0(p)&=&\frac{1}{2}\langle p'^2-\frac{(\textbf{p}\cdot\textbf{p}')^2}{\textbf{p}^2}\rangle
\nonumber\\
B_1(p)&=&\frac{1}{2}\langle 
\frac{(\textbf{p}\cdot\textbf{p}')^2}{\textbf{p}^2}-2\textbf{p}\cdot\textbf{p}'+\textbf{p}^2
\rangle \ .
\end{eqnarray}
Using the expression for $w(\textbf{p},\textbf{k}) $ in Eq.~(\ref{eq_wrate}) with the replacement 
$\textbf{k}-\textbf{p} \rightarrow \textbf{p}'$, and switching the integration variable to 
$ \textbf{p}' $, we express $ \langle X(\textbf{p}')\rangle$ in $T$-matrix form as 
\small
\begin{align}
\langle X(\textbf{p}')\rangle=&\sum_i\frac{1}{2 \varepsilon_Q(p)}\int 
\frac{d\omega'd\textbf{p}' }{(2\pi)^3 2\varepsilon_Q(p')}
\frac{d\nu d^3\textbf{q}}{(2\pi)^3 2\varepsilon_i(q)}
\frac{d\nu'd^3\textbf{q}'}{(2\pi)^3 2\varepsilon_i(q')}
\nonumber\\
&\times\delta^{(4)}\frac{(2\pi)^4}{d_Q}\sum_{a,l,s}|M|^2\rho_Q(\omega',p')\rho_i(\nu,q)
\rho_i(\nu',q')
\nonumber\\ 
&\times[1-n_Q(\omega')] n_i(\nu) [1\pm n_i(\nu')] X(\textbf{p}') \ .
\label{eq_offtrans}
\end{align}
\normalsize
The summation \(\sum_i\) is over all light flavors,  $i=u,\bar{u}, d, \bar{d}, s, \bar{s}, g$, 
where the light and strange quarks are assumed to have the same mass (which is a good approximation
in our context~\cite{Riek:2010py}). We include the relativistic phase space factor with the 
single-particle on-shell energy, denoted by $\varepsilon_{Q,i}(p)$. The heavy-light 
scattering matrix elements, $|M_{Qi}|^2$, in Eq.~(\ref{eq_offtrans}) are related to the $T$-matrix 
in the CM frame as
\begin{align}
& \sum_{a,l,s}\left|M^2\right| =16\varepsilon_Q(p_\text{cm})\varepsilon_i(p_\text{cm})
\varepsilon_Q(p_\text{cm}')\varepsilon_i(p_\text{cm}')] d^{Qi}_s
\nonumber\\
&\times\underset{a}{\sum }d^{Qi}_a\left| 4\pi\underset{l}{\sum}(2l+1)
T^{a,l}_{Q,i}(E_\text{cm},p_\text{cm},p_\text{cm}')P_l\left(\cos \theta _\text{cm}\right)\right|^2,
\label{eq_ampsq}	
\end{align}
where $T^{a,l}_{Qi}(E_\text{cm},p_\text{cm},p_\text{cm}')$ is calculated in the CM 
frame in all possible two-body color channels, $a$, and partial-wave channels, $l$. The CM 
energy, $E_\text{cm}$, incoming CM momentum $p_\text{cm}$, outgoing CM momentum $p_\text{cm}'$, 
and scattering angle $\cos \theta _\text{cm} $ are  expressed as functions of 
$E$, $\textbf{p}$, $\textbf{q}$, 
$\textbf{p}'$, $\textbf{q}'$, as discussed in App.~\ref{app_cm}. The two-body color/spin degeneracy 
factor is denoted by $d^{Qi}_{a,s}$, and the $P_l\left(\cos \theta_\text{cm}\right)$ are Legendre 
polynomials. The partial-wave summation is different from that employed in Eq.~(8) of 
Ref.~\cite{vanHees:2007me} (and in Ref.~\cite{Riek:2010py}), 
in that our expression (\ref{eq_ampsq}) includes the interference effects between different 
partial waves and an additional factor of $\pi$. We also carry the partial-wave expansion
to higher angular momenta of up to $l$=8 (compared to $l$=1 in 
Refs.~\cite{vanHees:2007me,Riek:2010py}), which turns out to be essential for the convergence 
of the high-momentum region of the transport coefficients. More explicitly, one can show that 
$\left|\sum_l (2l+1)c_l P_l(x)\right|^2=\sum_l(2l+1)b_lP_l(x)$, where each $b_l$ is a 
function of the $\{c_l\}$.
The final results for the friction coefficient using, \eg, the $U$ potential turn out to be 
within $\sim$20\% of the results of Ref.~\cite{Riek:2010py} based on the same lQCD 
free-energy data. This is a consequence of benchmarking the partial-wave expansion 
in both versions against the full perturbative-QCD (pQCD) results.

%%%%%%%%%%%%%%%%%%%%%%%%%%%%%%%%%%%%%%%%%%%%%%%%%%%%%%%%%%%%
\section{Charm Quark Transport Coefficients} %Charm-Quark Transport Coefficients
\label{sec_trans}
%%%%%%%%%%%%%%%%%%%%%%%%%%%%%%%%%%%%%%%%%%%%%%%%%%%%%%%%%%%%
In this section, we discuss the resulting charm-quark transport coefficients, focusing on the 
drag coefficient $A(p)$ which characterizes the thermal relaxation rate for the different input 
potentials.
We emphasize that the ``true" potentials $V_s$ and $V_w$ are part of a comprehensive many-body
set-up which encompasses the lQCD EoS and thus fully specifies the properties of thermal medium, 
\ie, the spectral functions (masses and widths) of the thermal partons that the 
heavy quark scatters off. This is not the case for the previously used potential ``proxies" 
$F$ and $U$, which have been applied within quasiparticle approximations for the QGP medium.    
Therefore, in Sec.~\ref{ssec_apquasi}, we first conduct baseline calculations for all four potentials, 
\{$ U $, $ F $, $ V_s $, $ V_w $\}, with thermal quasiparticle partons. 
In Sec.~\ref{ssec_apfull},  we employ the off-shell formalism outlined above to compute the transport 
coefficients for the potentials \{$ V_s $, $ V_w $\} in their accompanying bulk medium.
In Sec.~\ref{ssec_pvsnp} we scrutinize various nonperturbative effects (resummed vs. Born 
amplitudes, Coulomb vs. full calculations with string term, and on- vs. off-shell) to exhibit 
their quantitative role in the HQ transport.

%%%%%%%%%%%%%%%%%%%%%%%%%%%%%%%%%%%%%%%%%%%%%%%%%%%%%%%%%%%%%%%%%%%%%%%%%%%%%%
\subsection{Drag coefficients for different color forces in quasiparticle medium}
\label{ssec_apquasi}
\begin{figure} [!thb]
        \centering
        \includegraphics[width=0.99\columnwidth]{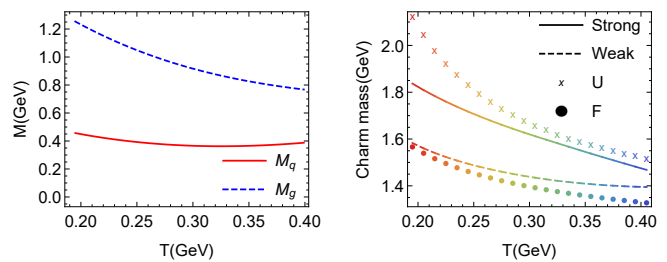}
        \caption{Light-parton (left) and charm-quark (right) masses for $V_s$ (solid lines),
$V_w$ (dashed lines), $U$ (crosses) and $F$ (dots) as used in the quasiparticle calculations 
leading to the results displayed in Figs.~\ref{fig_vufap}.}
        \label{fig_vufmass}
\end{figure}

%%%%%%%%%%%%%%%%%%%%%%%%%%%%%%%%%%%%%%%%%%%%%%%%%%%%%%%%%%%%%%%%%%%%%%%%%%%
We first restrict ourselves to the quasiparticle approximation for the QGP medium, \ie, 
the thermal-parton spectral functions in the expressions given in Sec.~\ref{sec_offtrans} 
are taken to be $\delta$-functions at their 
quasiparticle masses. The latter are chosen to be the same for all four potentials as shown in 
Fig.~\ref{fig_vufmass} left, obtained from a quasiparticle fit to the lQCD EoS using the Fock mass 
ansatz~\cite{Liu:2017qah} with $V_s$. The charm-quark masses, plotted in Fig.~\ref{fig_vufmass}
right, are taken to be $1.264+\Sigma(\infty;T)/2 $ where $\Sigma(\infty;T)$ 
denotes the infinite-distance limit of \{$U$, $F$, $V_s$, $V_w$\} as shown in Fig.~\ref{fig_pot}. 
Note that the light parton masses from the quasiparticle fit are different from the results 
extracted using the off-shell many-body calculations~\cite{Liu:2017qah}, while the charm-quark 
masses of \{$ V_s $, $ V_w $\} are taken from the corresponding potential. This setup allows for
an approximate ``apples-to-apples" comparison of how the different forces (or ``effective couplings") 
shown in Figs.~\ref{fig_vufforce}, \ref{fig_vasr} and \ref{fig_vasp} manifest themselves in 
the charm-quark transport coefficients.
\begin{figure} [!thb]
        \centering
        \includegraphics[width=0.99\columnwidth]{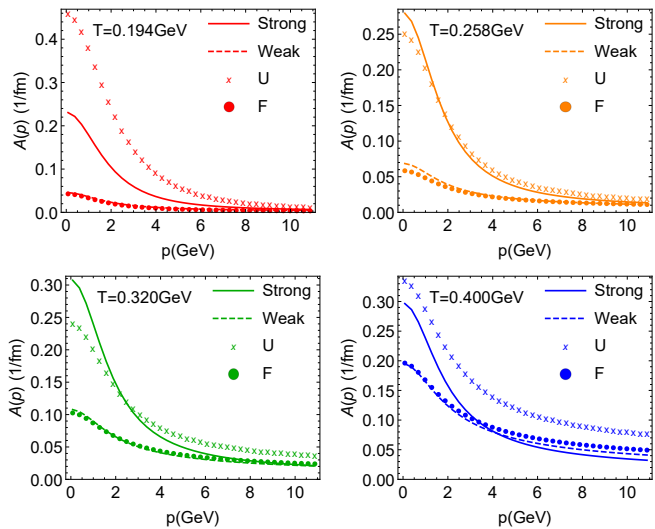}
        \caption{Friction coefficients for $V_s$ (solid line), $V_w$ (dashed line), $U$ (crosses)
and $F$ (dots) when using the on-shell Born diagrams in quasiparticle approximation.}
        \label{fig_vufapborn}
\end{figure}

We start with the case of using the Born approximation to calculate the friction 
coefficient, displayed in Fig.~\ref{fig_vufapborn} for the four potentials. 
The results for the WCS potential and the free energy closely agree across all temperatures 
and charm-quark momenta considered here. The friction
coefficient is much larger for the SCS potential and the internal energy, which are also rather 
close to each other except that the $U$-potential is about a factor 2 larger at the lowest 
temperature and at high momenta at the highest temperature. 

\begin{figure}[!t]
        \includegraphics[width=0.99\columnwidth]{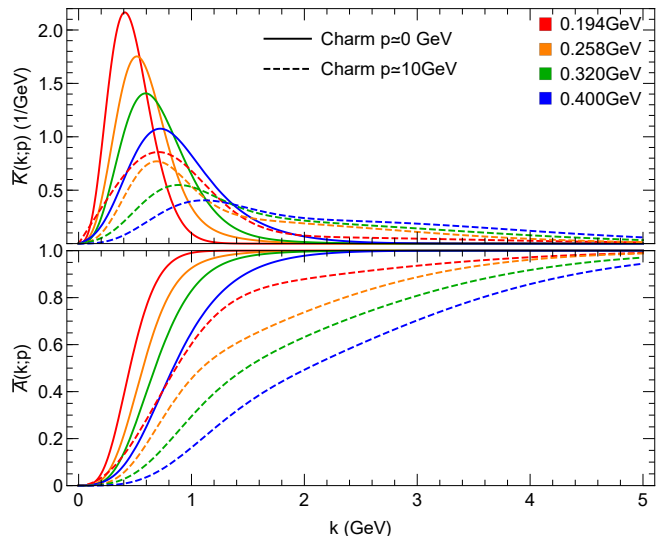}
    \caption{Differential CM momentum-transfer ``probability" distribution, 
$\bar{K}(k;p)$ (upper panel), for the friction coefficient from the SCS potential 
in Born approximation, and its cumulative (lower panel) for charm quarks at zero momentum 
($p$=0, solid lines) and $p$=10\,GeV (dashed lines) for different
temperatures (identified by the color code).}
\label{fig_ftoap}
\end{figure}

To better understand what the relevant momentum exchanges for the transport coefficients are, 
we divide up the phase space into shells of momentum transfer, $k dk$, where 
$k=|\vec p_\text{cm}- \vec p_\text{cm}'|$, and define a ``normalized" momentum-exchange 
density  
\begin{equation}
\bar{K}(k;p) dk \equiv A(p)^{-1} dA(k)  
\end{equation}
and a corresponding cumulative density 
\begin{equation}
\bar{A}(k;p) \equiv\int_{0}^{k}dk'\bar{K}(k';p) 
\end{equation}
of the friction coefficient, $A(p)$, defined such that $\bar{A}(p,k\to\infty)$=1.
These two quantities are plotted in Fig.~\ref{fig_ftoap} using the SCS potential, $V_s$ 
(still in quasiparticle and Born approximation).
For low-momentum charm quarks, most of the momentum transfers at low temperatures occur in a 
0.5\,GeV window around $k$=0.4\,GeV, corresponding to a relatively large force range of 
$\sim$1\,fm (recall the remark at the end of Sec.~\ref{sec_pot}). 
The peak position shifts to higher momentum transfer as temperature or HQ momentum increase, 
implying a transition from the long-range string force to a shorter-range Coulomb force.
This is due to a harder thermal phase space and the enhanced screening of the potential 
as temperature increases. 
For the $U$ potential, the effective coupling  at a momentum exchange of $\sim$0.5 GeV 
is about 50\% larger at the lowest temperature (recall upper left panel in Fig.~\ref{fig_vasp}),
leading to an approximately twice larger low-momentum friction coefficient in 
Fig.~\ref{fig_vufapborn}. A similar analysis applies to the other potentials. 

\begin{figure} [!t]
        \centering
        \includegraphics[width=0.99\columnwidth]{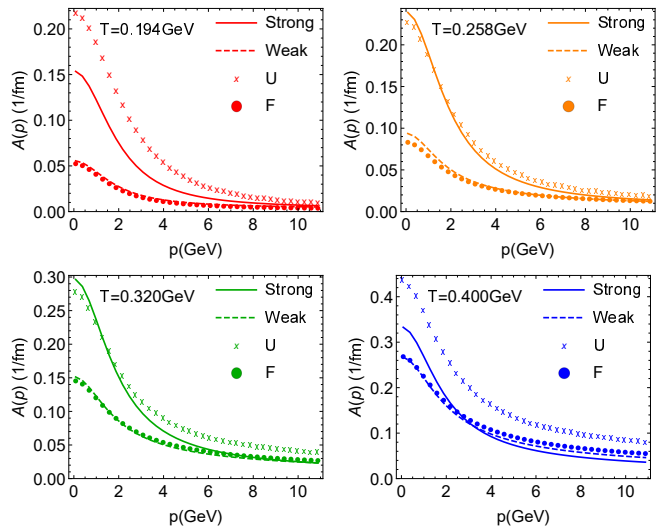}
        \caption{Quasiparticle friction coefficients for $V_s$ (solid line),
$V_w$ (dashed line), $U$ (crosses) and $F$ (dots).}
        \label{fig_vufap}
\end{figure}

In the next step, we compare the friction coefficients from the resummed $T$-matrix interactions
in Fig.~\ref{fig_vufap}, still using a quasiparticle QGP medium.
At low temperature and low momentum, the drag coefficients for $U$ and $V_s$ are {\em reduced} 
by a factor of 2 and 1.5, respectively, compared to the Born calculation. This is mainly because
the resummation converts the strongly attractive Born term into subthreshold resonance states
whose interaction strength is not accessible in 2$\to$2 on-shell scattering, while only a 
repulsive tail of the $T$-matrix remains in the on-shell phase space. However, for a less 
attractive potential which does not generate a strong bound state, which is the case for $F$ 
and $V_w$, the resummation generally enhances the Born result. On the other hand, at high 
momentum and high temperature, a closer agreement between the Born approximation and $T$-matrix 
results is found. 
%Not suprprisingly, the scaling behavior reflected by $\bar K$ at the Born level does not 
%straightforwardly Generalize to the (resummed) $T$-matrix interaction, especially at low momentum 
%and low temperature.

%%%%%%%%%%%%%%%%%%%%%%%%%%%%%%%%%%%%%%%%%%%%%%%%%%%%%%%%%%%%
\subsection{Transport coefficients with off-shell effects}
\label{ssec_apfull}
%%%%%%%%%%%%%%%%%%%%%%%%%%%%%%%%%%%%%%%%%%%%%%%%%%%%%%%%%%%
In the previous section we saw how in a strongly coupled medium the formation of bound states 
can lead to a marked {\em decrease} in the interaction strength when employing the 
quasiparticle approximation in two-body scattering. This should be considered as an artifact 
of an incompatible approximation.
In the presence of a large interaction strength, the single-particle spectral functions are
expected to become broad and/or develop collective modes below their nominal ``quasiparticle"
masses. In either case, phase space opens up below the quasiparticle two-body threshold
and allows for subthreshold resonance scattering. 
We now compute the charm-quark transport coefficients deploying the off-shell formalism described
in Sec.~\ref{sec_offtrans} to incorporate the quantum effects associated with subthreshold 
many-body interactions.
We focus on the results for the SCS and WCS as their selfconsistent solutions constructed in
Ref.~\cite{Liu:2017qah} specify the spectral functions of the thermal partons, while this
information is not available for $U$ nor $F$.

\begin{figure*} [!htb]
	\centering
	\includegraphics[width=2\columnwidth]{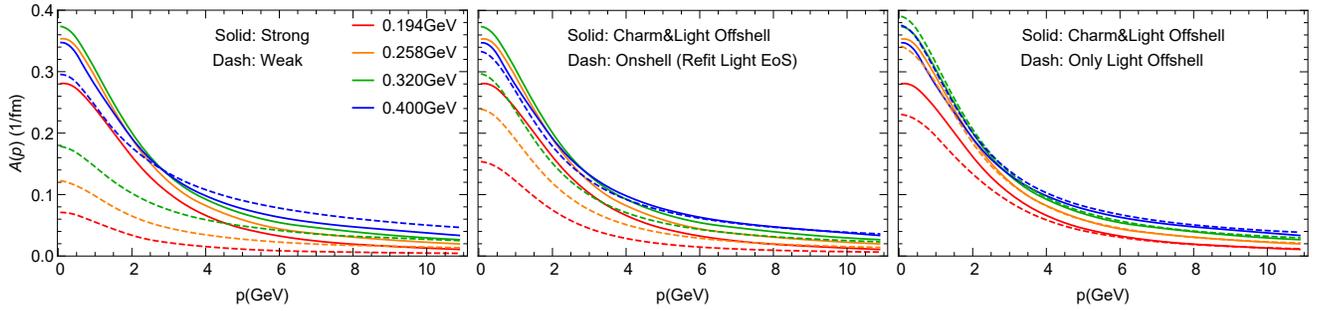}
	\caption{Charm-quark friction coefficients, $A(p)$, for the full off-shell calculations
(left) in the SCS (solid lines) and WCS (dashed lines), and comparing the full off-shell case 
for the SCS (solid lines) with one using the on-shell approximation for both thermal partons and 
the outgoing charm quark (dashed lines; middle panel) or for the outgoing charm quark only (dashed 
lines in right panel).} 
	\label{fig_Ap}
\end{figure*}
\begin{figure} [htb]
	\centering
	\includegraphics[width=0.8\columnwidth]{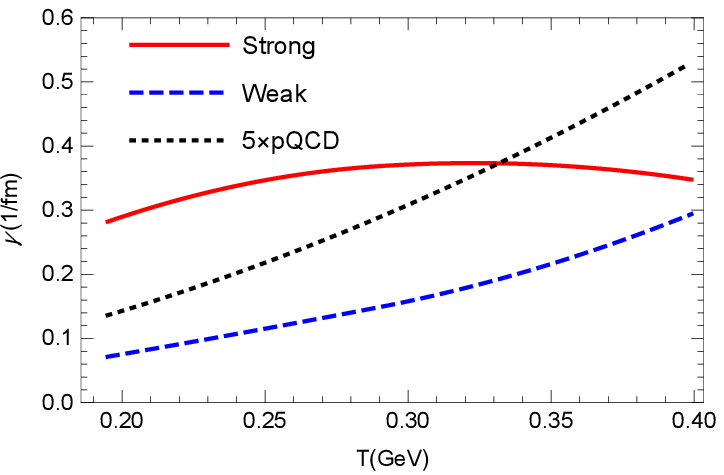}
	\includegraphics[width=0.8\columnwidth]{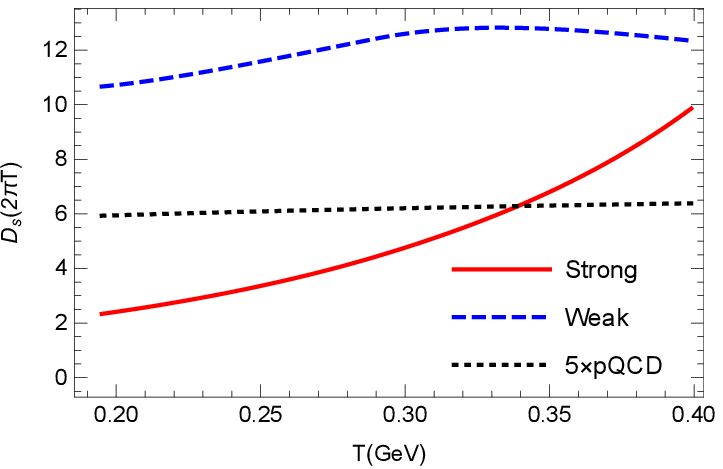}
	\vspace{-0.15cm}
	\vspace{-0.15cm}
	\caption{Temperature dependence of the zero-momentum relaxation rate, 
		$\gamma$ (upper), and the spatial diffusion coefficient, 
		$D_s=T/(\gamma M_c)$ (lower, in units of the thermal wave length $D_s(2\pi T)$). 
		The pQCD results uses $\alpha_s=$0.4 and a factor of 5.}
	\label{fig_gammads}
\end{figure}
The pertinent charm-quark friction coefficients are compiled in Fig.~\ref{fig_Ap}. 
The full results displayed in the upper left panel show that for small momenta and small
temperatures the relaxation rate is about four times larger for the SCS than for the WCS, 
while with increasing momentum and temperature they approach each other.
The key reason for the large enhancement at low momentum and temperature is the remnant 
of the long-range confining force, as discussed in the context of Figs.~\ref{fig_pot}, 
\ref{fig_vufforce}, \ref{fig_vasr} and \ref{fig_vasp}.  
At higher temperatures, the confining potential is largely screened, and the larger thermal 
parton momenta probe the force at shorter distances. Since the short-range Coulomb force is 
quite similar for the WCS and SCS, the difference between $ A_s(p) $ and $ A_w(p) $ is reduced 
(in the fits of Ref.~\cite{Liu:2017qah} the screening of the Coulomb interaction is slightly 
weaker in the WCS than in the SCS, causing $ A_w(p) $ to exceed $ A_s(p) $ at high momenta and 
at the highest temperature where the confining interaction has nearly vanished).

The off-shell effects in the SCS scenario are illustrated in the middle and right panel of 
Fig.~\ref{fig_Ap}, where we have switched them off for either both thermal partons and
the outgoing charm quark (middle panel) or only for the outgoing charm quark (right panel). 
In the former case, we have re-adjusted (\ie, decreased) 
the thermal parton masses to ensure compatibility with the lQCD EoS.  
%This is a genuine many-body effect, since the width of the parton is caused by many-body collisions
%---as illustrated in Fig.~\ref{fig_manybody}. 
We find that the quantum effects almost double the transport coefficients in the small-momentum 
and low-temperature region: the broadening of the thermal spectral functions allows to probe 
off-shell energies in the $T$-matrix where scattering through a (broad) bound state becomes 
possible. This confirms, in a more rigorous treatment, the original findings of 
Refs.~\cite{vanHees:2004gq,vanHees:2007me}, where near threshold resonances were 
put forward to solve the heavy-flavor puzzle in Au-Au collisions at RHIC~\cite{Adare:2006nq}. 
A more moderate but still significant effect arises from the non-trivial spectral function of the
outgoing charm quark. Switching back to a $\delta$-function reduces the low-momentum low-temperature 
relaxation rate by almost 20\%, cf.~right panel of Fig.~\ref{fig_Ap}. Once the resonance states 
are close to threshold (or have melted) so that the on-shell treatment can already access the
main scattering strength, the off-shell treatment does not provide a significant enhancement.
For the WCS, the results from the full off-shell case generally agree well with the 
results from the quasiparticle case (not shown), since the widths of spectral functions are small. 
At high momentum, the HQ drag coefficients are dominated by the Coulomb term, augmented 
by relativistic (magnetic) corrections (Breit enhancement), while the scalar vertex assumed 
for the string interaction suppresses its high-momentum contribution. 
Therefore, the off-shell case approaches the quasiparticle case: the spectral functions become  
more quasiparticle like, and the typical CM energy in the $T$-matrix becomes larger so that even 
off-shell effects do not significantly probe the subthreshold resonances anymore.

In Fig.~\ref{fig_gammads} we summarize the temperature dependence of the zero-momentum 
relaxation rate, $\gamma=A(p\rightarrow 0)$, and the dimensionless spatial diffusion 
coefficient, $D_s(2\pi T)=(2\pi T^2)/(m_c \gamma_c)$, for the WCS and SCS. As a reference, 
we also show a perturbative 
QCD (pQCD) Born result (using $\alpha_s$=0.4 in a quasiparticle QGP with Debye and thermal 
parton masses of $gT$, and a constant charm-quark mass of 1.5\,GeV) upscaled by a factor of 
5 (as recently used as a benchmark scenario in Ref.~\cite{Rapp:2018qla}). The temperature 
behavior of the relaxation rates and spatial diffusion coefficients for the WCS is similar 
to the pQCD*5 scenario, wherein $\gamma$ increases monotonically with temperature and $D_s(2\pi T)$ 
is essentially constant, similar to what one would expect from a dimensionless theory. 
For the SCS, on the other hand, $\gamma$ exhibits a rather flat behavior with temperature where 
the increasing density of the thermal scatterers is essentially compensated by the decreasing
interaction strength. Consequently, $D_s(2\pi T)$ increases with temperature by about a factor 
5 over the considered temperature range of $T$=0.2-0.4\,GeV; the extra dimensionful quantity is 
brought in by the nonperturbative string tension. Also note that the SCS diffusion coefficient 
differs from the ``bare" pQCD interaction by a factor of almost 15 at low temperature.

%%%%%%%%%%%%%%%%%%%%%%%%%%%%%%%%%%%%%%%%%%%%%%%%%%%%%%%%%%%%%%
\subsection{Scrutinizing Nonperturbative Effects}
\label{ssec_pvsnp}
\begin{figure} [!hbt]
	\centering
	\includegraphics[width=0.99\columnwidth]{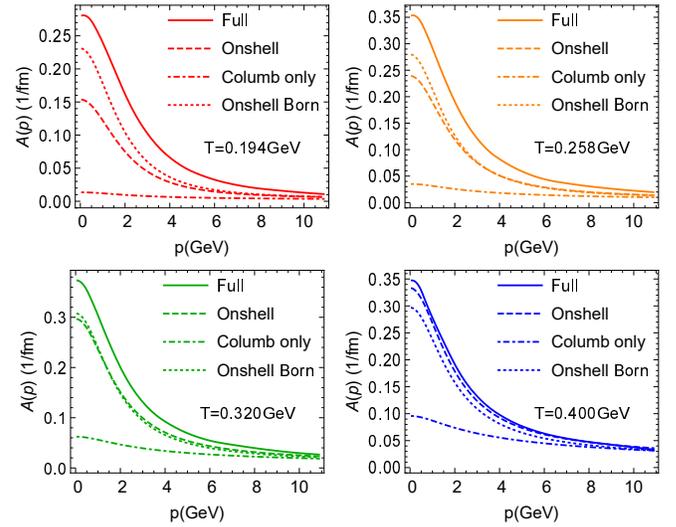}
	\caption{Comparison of the effects of different ingredients on the HQ transport 
coefficients in the SCS as a function of charm-quark momentum, at 4 different temperatures. 
Solid lines: full results; dashed lines: using the on-shell approximation for the thermal 
partons and outgoing charm-quark; dash-dotted line: on-shell results using only the Coulomb 
term in the potential;  dotted lines: using the Born and quasiparticle approximation (including 
the confining potential).}
	\label{fig_Appert}
\end{figure}

%%%%%%%%%%%%%%%%%%%%%%%%%%%%%%%%%%%%%%%%%%%%%%%%%%%%%%%%%%%%%%
In the calculation of the transport coefficients, there are at least three nonperturbative 
components: (1)~the string interaction in the potential; 
(2)~the resummation of the $T$-matrix possibly leading to the resonance formation; 
(3)~off-shell effects from the large widths of the partons. 
Here, we reassess these effects relative to the full calculation of the friction
coefficient within the SCS in Fig.~\ref{fig_Appert}, using the thermal parton and 
charm-quark masses shown in Fig.~\ref{fig_vufmass}.

When switching off the string interaction in the potential (and neglecting off-shell effects,
which play a negligible role in this scenario), the pertinent $T$-matrix results for the 
friction coefficient (labeled ``Coulomb-only'' in Fig.~\ref{fig_Appert}) are much reduced 
compared to the full results at low momentum, close to a factor of 15 at low temperature, and 
still by a factor of $\sim$3 at $T$=0.4\,GeV. At charm-quark momenta of $p$=10\,GeV, the 
reduction is still significant at low $T$ (indicating a non-negligible portion of soft 
interactions driven by the string term), but has essentially ceased at $T$=0.4\,GeV.   
Therefore, perturbative (elastic) calculations of $A(p)$ that do not account for remnants of
the confining term are not be reliable at low temperatures even at momenta of \(p=10\)\,GeV.
The ``on-shell" results with the full interaction, already shown in the previous section, fall 
below the full results by almost 50\% at low $T$ and nearly uniform in 3-momentum from 0 to 
10\,GeV.  This implies that even at $p$=10\,GeV, the soft off-shell effects (making accessible 
the subthreshold resonances) are significant, although in practice one expects radiative processes 
to become dominant at these momenta. The difference between full and on-shell calculations is 
essentially gone at $T$=0.4\,GeV (resonance structures have ceased), again basically across the 
entire momentum range. Finally, the ``on-shell Born" results are surprisingly close to full 
results within a few 10's of percent. This is, however, a highly deceptive result: if we 
include the second Born term in the $T$-matrix, the friction coefficient is up to 5 times larger 
at low momentum and low temperature, signaling an uncontrolled convergence property of the 
perturbative series at low momentum, very similar to the findings in Ref.~\cite{CaronHuot:2007gq}. 
This is another reminder that a proper resummation in the nonperturbative region is mandatory. 
%However, in nonrelativistic Coulomb 
%scattering, with exact  $ 1/r $ as potential, the full resumed solution is the same as Born 
%approximation and classical Rutherford scattering at arbitrary large  coupling. This may be 
%one reason why the full results are close to the Born approximation at strong coupling while 
%the 2.~order blows up. Also, this means that the Born term giving a reasonable result is not 
%implying that the perturbative approximation works since the reason behind it may be fully 
%nonperturbative. 
Figure~\ref{fig_Appert} furthermore shows that the ``on-shell Born'' and ``on-shell'' curves 
approach each other at high momentum. Still, the results for the second Born order at high 
momentum and low temperature double the first-order result, \ie, the convergence of the 
perturbative series is still not good (due to the presence of the string term). This situation
improves at higher temperature: at $ T=$0.4~GeV, the second Born contribution is only by a factor 
1.8 (1.6) larger than the Born contribution at low (high) momentum.

%The reason for the off-shell effects has large impact on transport coefficients  

%%%%%%%%%%%%%%%%%%%%%%%%%%%%%%%%%%%%%%%%%%%%%%%%%%%%%%%%%%%%
\section{Charm-Quark Langevin Simulations in Heavy-Ion Collisions} 
\label{sec_urhic}
\begin{figure} [!t]
	\centering
	\includegraphics[width=1\columnwidth]{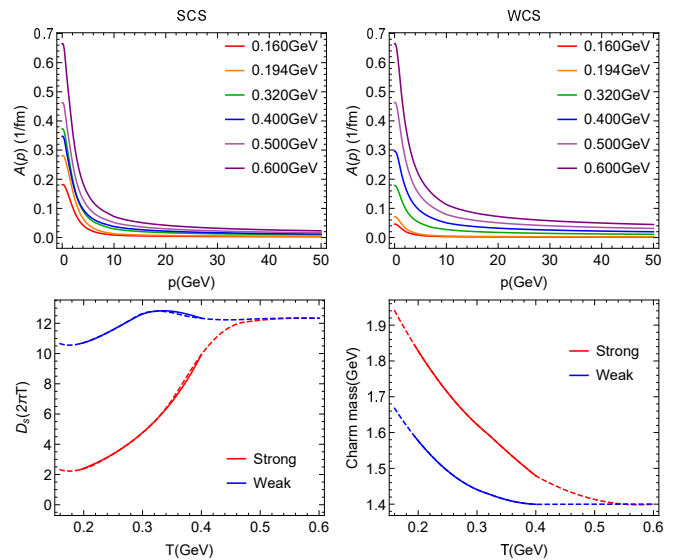}
	\caption{Extrapolation results for $ D_s(2\pi T) $, $ M_c $, $ A_s(p) $ and $ A_w(p) $.}
	\label{fig_extra}
\end{figure}

%%%%%%%%%%%%%%%%%%%%%%%%%%%%%%%%%%%%%%%%%%%%%%%%%%%%%%%%%%%%
In this section we implement the transport coefficients following from the selfconsistent 
WCS and SCS, as well as from the $U$-potential proxy with quasiparticle QGP medium, 
into Langevin simulations of charm quarks in URHICs as described in Ref.~\cite{He:2011qa}.
% obtain the D meson observables~\cite{He:2017}.
%\footnote{Min He did the simulation with our new transport coefficients} 
As our current calculations are limited to temperatures $T$=0.194-0.4\,GeV and momenta 
$p$=0-10\,GeV, an extrapolation is required to cover the ranges needed in the Langevin 
approach to heavy-ion collisions at the LHC.
%Since the evaluation for $A(p)$ at 
Since the $p$-dependence of the quasiparticle results is similar to the full results 
at high momentum (as discussed in the previous section), we extrapolate $A(p)$ to higher momenta 
using the quasiparticle results augmented by a $p$-independent $K$ factor to smoothly connect 
them at $p$=10\,GeV. For the extrapolation to lower and higher temperatures, we first extrapolate 
$D_s(2 \pi T)$ and $m_c$  as shown in the lower two panels of Fig.~\ref{fig_extra}. Then, we use 
$A(p=0;T)=T/(D_s m_c)$ and take
the momentum dependence of $A(p;T)$ to be the same as for $A(p;T)$ at $T$=0.194(0.4)\,GeV 
for low (high) temperature, as shown in the upper two panels of Fig.~\ref{fig_extra}.

The transport coefficients are utilized within the Langevin equations
\begin{align}
&d\textbf{x}=\frac{\textbf{p}}{\varepsilon_c(p)} dt\\
&d\textbf{p}=\Gamma(p)\,\textbf{p}dt+\sqrt{2 dt D(p)}\bm{\rho}  \ ,
\end{align}
where the relaxation rate, $\Gamma(p)$, and the momentum diffusion coefficient, $D(p)$, are 
taken to be $\Gamma(p)=A(p)$ and $D(p)=B_0(p)=B_1(p)=T\varepsilon_c(p)\Gamma(p)$, and 
$ \bm{\rho} $ is a random number determined from the Gaussian distribution function 
$ P(\bm{\rho})=(2\pi)^{-3/2}e^{-\bm{\rho}/2} $. Using the Langevin equations, we simulate 
Brownian motion of charm quarks in a background medium provided by an ideal hydrodynamic 
evolution of the QGP fireball in URHICs at RHIC and the LHC. For definiteness, we choose 
semicentral Pb-Pb collisions at CM energy $\sqrt{s_{\rm NN}}$=5.02\,TeV, at a fixed impact 
parameter representing the 20-40\% centrality class. 

\begin{figure} [!tb]
                \includegraphics[width=0.99\columnwidth]{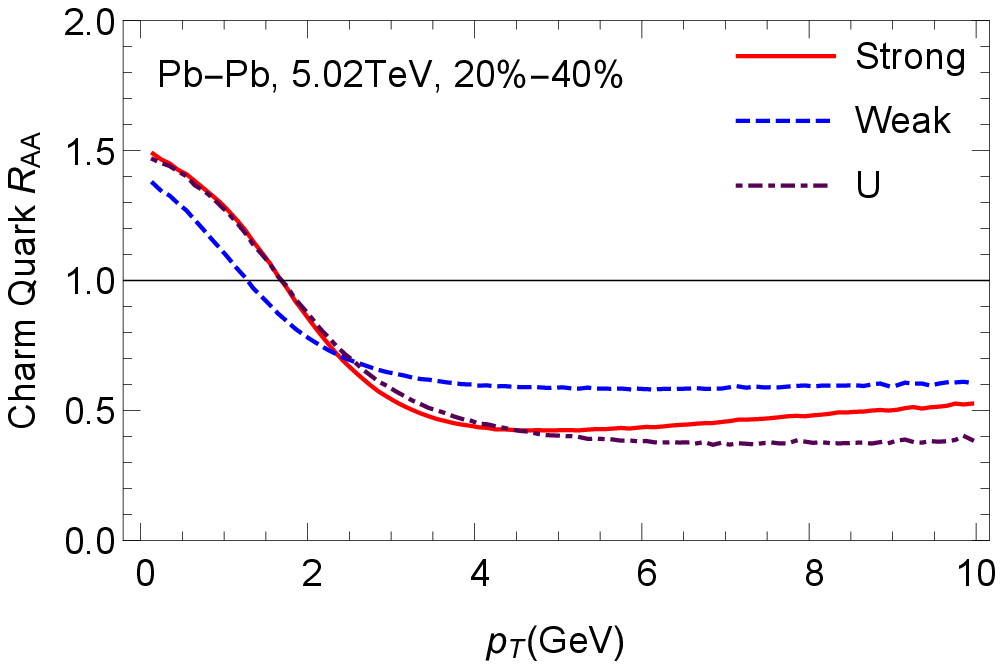}
        \includegraphics[width=0.99\columnwidth]{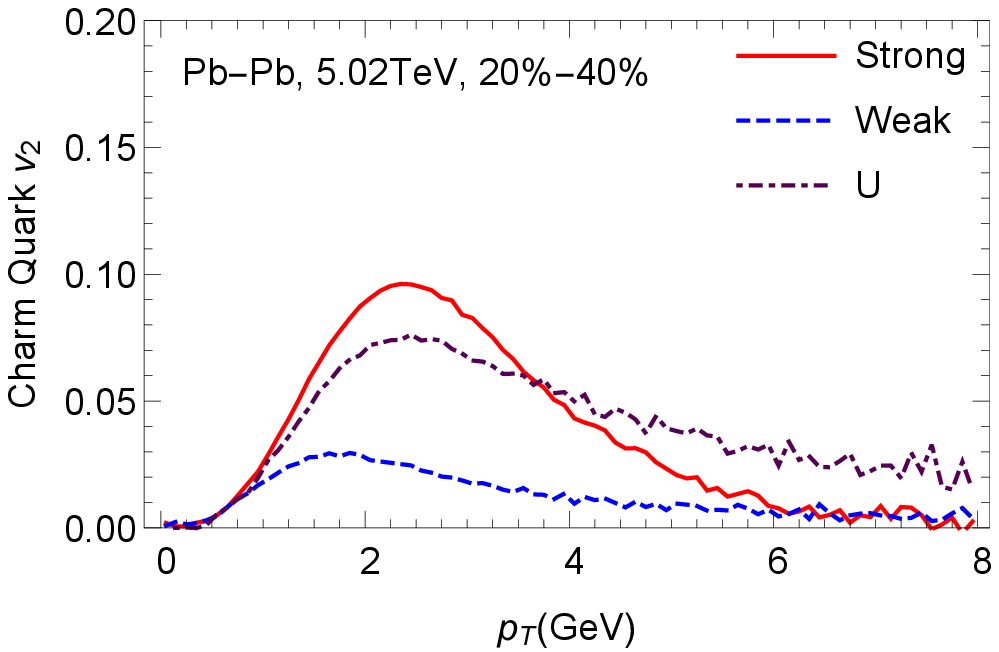}
        \caption{The $R_{\rm AA}$ (upper panel) and $v_2$ (lower panel) of charm quarks
calculated from the $T$-matrix interactions with 3 different potentials using
relativistic Langevin simulations in a hydrodynamic fireball evolution for semicentral
Pb-Pb collisions at the LHC.}
        \label{fig_c-RAA-v2}
\end{figure}

Figure~\ref{fig_c-RAA-v2} summarizes the nuclear modification factor, $R_{\rm AA}$, and elliptic
flow, $v_2$, of charm quarks at the end of the QGP evolution, taken at $T_{\rm pc}$=170\,MeV, for 
the three potentials under investigation. The $R_{\rm AA}$ shows the standard feature of softening
the initial charm-quark spectra, but only exhibits a modest sensitivity to the underlying
potential. This reiterates the finding~\cite{Rapp:2008qc} that the main effects determining
the charm-quark $R_{\rm AA}$ occur early in the evolution where the difference between the
potentials is small. This is quite different for the elliptic flow~\cite{Rapp:2008qc}, 
which requires several fm/$c$ to build up in the expanding fireball. At that point the
difference in the underlying potential scenarios becomes maximal, and, consequently, 
the low-$p_t$ elliptic flow of charm quarks provides a direct gauge of the coupling
strength in the later stages of the QGP evolution. More quantitatively, the largest value
of the $v_2$ is generated within the SCS reaching near 10\%, more than a factor of 3 larger 
than in the WCS. It also exceeds the maximum value attained with the $U$-potential proxy     
by about 20\%, indicating that low-$p_t$ elliptic flow of charm quarks is rather sensitive
to the long-distance behavior of the in-medium potential, and thus an excellent measure 
of the spatial diffusion coefficient. Note that a charm-quark momentum of $p_t$=2\,GeV
corresponds to a velocity of about 0.74$c$, not much larger than the (surface) flow 
velocities reached in the fireball expansion at the end of the QGP phase. At higher $p_t$, 
above $\sim$4\,GeV, the intermediate-distance strength is largest in the $U$-potential
and leads to significantly larger $v_2$ values than obtained for $V_s$ and $V_w$. 

\begin{figure} [!tb]
        \centering
        \includegraphics[width=0.99\columnwidth]{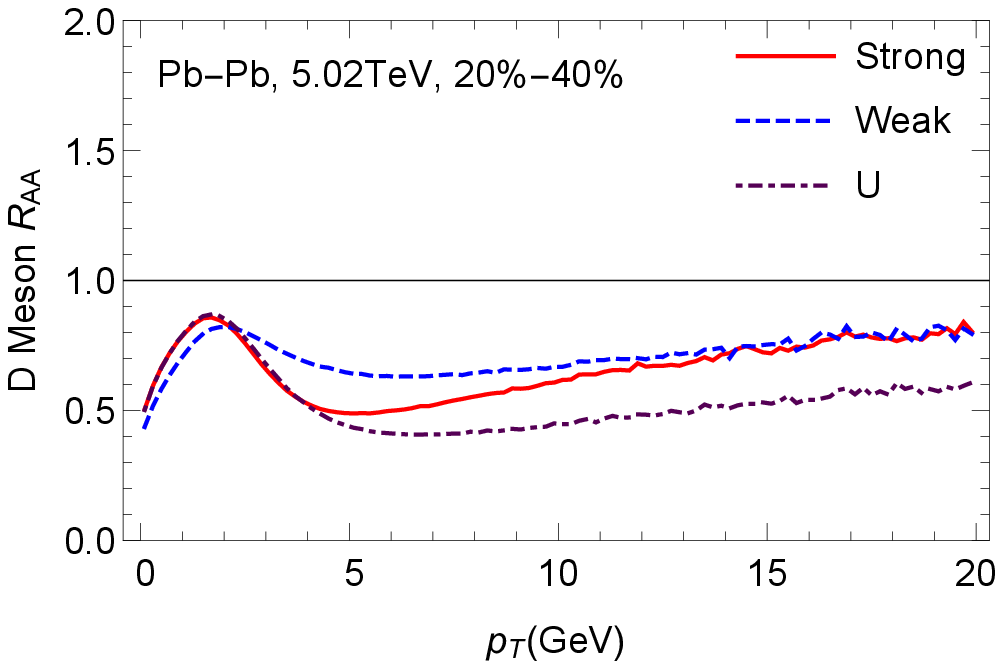}
        \includegraphics[width=0.99\columnwidth]{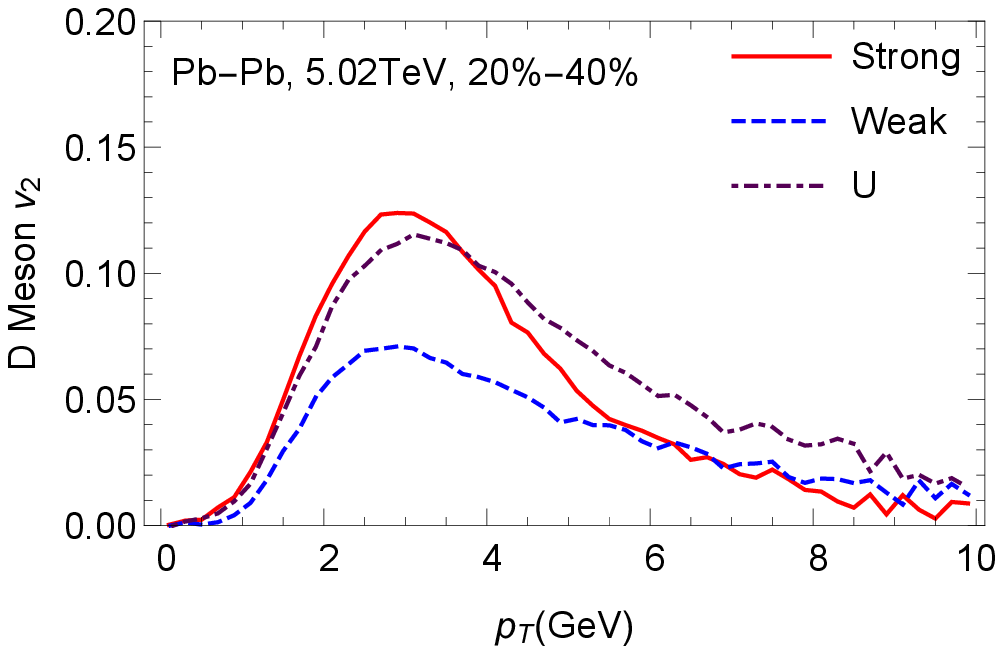}
        \caption{Comparison of the calculated $D$-meson $R_{\rm AA}$ (upper panel) and $v_2$
(lower panel) obtained from applying a recombination-fragmentation model to hadronize the
charm-quark spectra plotted in Fig.~\ref{fig_c-RAA-v2}.}
        \label{fig_D-RAA-v2}
\end{figure}

To make contact with experiment, we proceed to calculate $D$-meson observables. 
As the fireball medium approaches the pseudo-critical temperature, charm quarks are 
hadronized into $D$ mesons through either recombination with surrounding light quarks from 
the hydrodynamic medium (pre-dominantly at low $p_t$)~\cite{Ravagli:2007xx} or independent 
fragmentation (we also account for a $\sim$20\% ($p_t$-dependent) reduction in the $D$-meson 
yields due to shadowing and ``chemistry effects" where charm quarks hadronize into other 
hadrons like $D_s$ and $\Lambda_c$ at a higher fraction than in proton-proton collisions). 
We finally carry out the $D$-meson diffusion in the hadronic phase. The resulting $D$-meson 
$R_{\rm AA}$ and $v_2$ are shown in Fig.~\ref{fig_D-RAA-v2}. Recombination effectively acts 
as another interaction between charm quarks and the medium, driving the $D$-meson spectra 
closer to equilibrium~\cite{He:2011qa}.
This produces a characteristic flow ``bump" in the $R_{\rm AA}$ at a $p_T$ reflecting 
the velocity of low-momentum $D$-mesons embedded in the flowing hydrodynamic medium. At 
high $p_T$, fragmentation takes over, and the $D$-meson $R_{\rm AA}$ tends toward that 
of the charm quark 
(modulo further suppression due to $D$-meson interactions in the hadronic phase). 
Other than the flow bump, the qualitative features of the charm-quark spectra relating
to the different potentials are preserved at the $D$-meson level. However, the 
discrimination power is reduced, especially for the maximum value of the $v_2$, 
which is now quite similar for the SCS and the $U$-potential, while the $v_2$ of the
WCS is only a factor 2 below the former two. This is because recombination plus hadronic
diffusion together add a roughly equal amount of $v_2$ in the 3 potential scenarios when 
going from charm-quark to $D$-meson spectra. To some extent this is an artifact of applying 
the same coalescence model to all three scenarios. In
reality, the coalescence probability should be smaller in the WCS compared to the SCS, 
since the $D$-meson resonance strength, which is the microscopic mediator of the recombination
process, is weaker in the WCS than in the SCS and thus should lead to a smaller increment
in $v_2$ in the former compared to the latter. While the resonance recombination 
model~\cite{Ravagli:2007xx} (as employed here) in principle encodes this mechanism, its 
implementation in the current calculation does not account for this difference. 
These considerations reiterate the importance of a recombination model that is consistent 
with the microscopic interactions driving the diffusion process in the the vicinity 
of $T_{\rm pc}$.  

Recalling the experimental results~\cite{ALICE:2012ab,Abelev:2013lca,Sirunyan:2017xss}, which 
report maximal $v_2$ values of $D$-mesons in 30-50\% central collision of ca.~17$\pm$2\%, the 
SCS scenario is not far below, but the WCS and also the free-energy potential (not shown here) 
are strongly disfavored as their interaction strength is too small. At higher momenta, 
$p_T\gtsim5$\,GeV, both WCS and SCS produce too little suppression and too little $v_2$ (even
the $U$ potential did not supply enough suppression in comparisons to central Pb-Pb data). 
This is, however, expected, since radiative processes 
%\cite{Mustafa:2004dr,Gossiaux:2010yx,Das:2010tj,Abir:2012pu,Cao:2013ita} 
have not been systematically included yet (some are encoded through the in-medium selfenergies 
of the heavy and light quarks in the $T$-matrices), see, \eg, Ref.~\cite{Rapp:2018qla} for a 
recent discussion and references. 
Such processes may also help to reduce the milder discrepancies at lower $p_T$. 
It remains to be seen whether contributions beyond the potential approximation might be 
helpful in generation additional interaction strength. 
From a more practical perspective, fluctuating initial conditions in the hydrodynamic evolution 
are conceivable for producing an enhancement in the $v_2$ over the results from smooth 
initial conditions~\cite{Nahrgang:2014vza,Noronha-Hostler:2016eow,Prado:2016szr,Cao:2017umt} 
as employed in the ideal hydrodynamic evolution used here.

%%%%%%%%%%%%%%%%%%%%%%%%%%%%%%%%%%%%%%%%%%%%%%%%%%%
%\subsection{Comparison to RHIC Data}
%\label{ssec_obs2}
%%%%%%%%%%%%%%%%%%%%%%%%%%%%%%%%%%%%%%%%%%%%%%%%%%%

%\FloatBarrier
%%%%%%%%%%%%%%%%%%%%%%%%%%%%%%%%%%%%%%%%%%%%%%%%%%%%%%%%%%%%%%%%
\section{Conclusions and Perspectives}
\label{sec_concl}
%%%%%%%%%%%%%%%%%%%%%%%%%%%%%%%%%%%%%%%%%%%%%%%%%%%%%%%%%%%%%%%%
In an attempt to establish connections between heavy-flavor phenomenology in heavy-ion
collisions and the microscopic interactions driving the diffusion of heavy quarks through the 
QGP formed in these reactions, we have employed a range of underlying two-body interaction 
potentials to compute the heavy-light $T$-matrices and pertinent HQ transport coefficients. 
Specifically, we have investigated two potentials recently constructed to satisfy constraints 
from lQCD for HQ free and internal energies, quarkonium correlators and the QGP EoS, as 
well as the free and internal energies, which have been used previously as potential proxies. 
We have first analyzed the corresponding forces, in particular their typical ranges in both 
coordinate and momentum space. As expected, the $U$-potential yields the largest force strength, 
realized at intermediate distances, while the strongly coupled $T$-matrix solution develops 
a smaller force but of longer range; in both cases the remnants of the confining force in the 
QGP play a key role in generating nonperturbative interaction strength, operative for temperatures 
of up to about 2.5\,$T_{\rm pc}$. The weakly coupled solution and the free energy have very 
similar forces, but are further reduced in strength and of much shorter range than the internal 
energy and the strongly coupled solution.
We then derived a transport equation including quantum many-body (off-shell) effects, to account
for the broad spectral functions of the thermal medium partons characterizing, in particular, the 
SCS of the $T$-matrix solution. These off-shell effects are instrumental in enabling the diffusing 
heavy quarks to probe the interaction strength of the broad subthreshold two-body resonances in the 
heavy-light scattering amplitudes. As a somewhat surprising result, the SCS potential develops the 
largest thermal relaxation rate for low-momentum charm quarks among all four potentials, while the 
$U$-potential's rate is strongest at intermediate and large momenta.   
Implementing these potentials into relativistic Langevin simulations revealed the SCS potential
to develop the largest peak value of the charm-quark $v_2$, about 20\% above the $U$-potential 
and a factor of 3 larger than the WCS potential (or free energy). Computing pertinent $D$-meson 
observables and benchmarking them against experimental data at the LHC rules out the WCS and 
free energy as viable potentials for HQ interactions in the QGP. Even the SCS potential falls 
slightly short of accounting for the existing low-momentum $v_2$ data at the LHC.    
These findings imply that charm quarks acquire collisional widths of 0.5-1\,GeV in the QGP near 
$T_{\rm pc}$, and consequently low-momentum light partons are likely dissolved in this regime,
\ie, soft excitations in the QGP near $T_{\rm pc}$ do not support parton quasiparticles; at the 
same time, broad hadronic resonances emerge and act as mediators of the nonperturbative interaction 
strength.   

Among the challenges that remain in the HQ sector, from a microscopic point of view, are to 
account for the missing 20\% in the elliptic flow of $D$-mesons as observed at the LHC, and to 
incorporate gluon radiation in a strongly-coupled framework. The latter will be essential to
increase the high-$p_T$ suppression and $v_2$, whereas genuine 3-body scattering, retardation 
effects, improvements in the coalescence mechanism and/or the hadronic diffusion, as well as 
features of the bulk evolution not captured by the ideal-hydro model employed here, could 
augment the $v_2$ at low-$p_T$.    
Work on several aspects of the above has already been done by various groups and/or is in 
progress, and efforts to combine them are ongoing~\cite{Rapp:2018qla} and expected to reveal 
further insights in due course.

\acknowledgments
This work was supported by the U.S.~National Science Foundation (NSF) through
grant PHY-1614484, by the A.-v.-Humboldt Foundation, and by the NSFC grant 11675079.

\appendix

%%%%%%%%%%%%%%%%%%%%%%%%%%%%%%%%%%%%%%%%%%%%
\section{Center-of-Mass Transformation}
\label{app_cm}
%%%%%%%%%%%%%%%%%%%%%%%%%%%%%%%%%%%%%%%%%%%%
In this appendix, we provide details on the CM transformation implemented in this work. We first
discuss the CM transformation in a nonrelativistic system, followed by the relativistic case. 

The nonrelativistic $T$-matrix can be expressed as
\begin{align}
&T(E,\textbf{P},\textbf{p}_1,\textbf{p}_1')=V(\textbf{p}_1-\textbf{p}_1')+
\nonumber\\
&\int_{-\infty}^{\infty}\frac{d^3\textbf{k}_1}{(2\pi)^3}
V(\textbf{p}_1-\textbf{k}_1)G^{(2)}_{(0)}(E,\textbf{k}_1,\textbf{P}-\textbf{k}_1)
T(E,\textbf{P},\textbf{k}_1,\textbf{p}_1'). \label{eq_BESNRwithP}
\end{align}
where the total 3-momentum and energy are $\textbf{P}=\textbf{p}_1+\textbf{p}_2$  and
$E=\omega_1+\omega_2+i\epsilon$, respectively. In a nonrelativistic system, the two-body 
propagator reads
\begin{align}
&G^{(2)}_{(0)}(E,\textbf{k}_1,\textbf{P}-\textbf{k}_1)=\frac{1}{E-\frac{\textbf{k}_1^2}{2M_1}-
\frac{(\textbf{P}-\textbf{k}_1)^2}{2M_2}} \  ,
\label{eq_G2NR}
\end{align}
and the CM transformation can be expressed as
\begin{align}
&\textbf{v}_\text{cm}=\frac{\textbf{P}}{M_{\rm tot}} \ ,
\\
&\textbf{p}_{{\rm cm}}=\textbf{p}-M_i\textbf{v}_\text{cm} \ ,
\end{align}
where $\textbf{p}$ (and the corresponding $\textbf{p}_{{\rm cm}}$) is a generic notation 
for $\textbf{p}_{1,2}, \textbf{p}_{1,2}', \textbf{k}_{1}$, etc., and $M_{\rm tot}=M_1+M_2$
is the total mass. 
Therefore, the transformations for the momenta are 
\(\textbf{p}_1=\textbf{p}_\text{cm}+M_1\textbf{v}_\text{cm},
\textbf{p}_1'=\textbf{p}_\text{cm}'+M_1\textbf{v}_\text{cm},
\textbf{k}_1=\textbf{k}_\text{cm}+M_1\textbf{v}_\text{cm}\). 
Substituting these into Eq.\,(\ref{eq_BESNRwithP}) and
noting that  \(V(\textbf{p}_1-\textbf{p}_1')=V(\textbf{p}_\text{cm}-\textbf{p}_\text{cm}')\), 
we obtain an equivalent equation that only depends on \(E\) and \(\textbf{P}\) implicitly 
through \(E_\text{cm}=E-\textbf{P}^2/(2M_{\rm tot})\),
\begin{align}
&T(E_\text{cm},\textbf{p}_\text{cm},\textbf{p}_\text{cm}')=
V(\textbf{p}_\text{cm}-\textbf{p}_\text{cm}')
+\int_{-\infty}^{\infty}\frac{d^3\textbf{k}_\text{cm}}{(2\pi)^3}
\nonumber\\
&V(\textbf{p}_\text{cm}-\textbf{k}_\text{cm})\frac{1}{E_\text{cm}
-\frac{(\textbf{k}_\text{cm})^2}{2\mu}}T(E_\text{cm},\textbf{k}_\text{cm},\textbf{p}_\text{cm}') \ ,
\label{eq_NRTmCM}
\end{align}
with the reduced mass $\mu=M_1M_2/M_{\rm tot}$. 
The solution to the original equation~(\ref{eq_BESNRwithP}) is calculated using the reverse 
CM transformation,
\begin{align}
&\textbf{p}_\text{cm}=\textbf{p}_1-\frac{M_1\textbf{P}}{M_{\rm tot}}=\frac{\textbf{p}_1M_2-\textbf{p}_1M_1}{M_{\rm tot}}
\nonumber\\
&\textbf{p}_\text{cm}'=\textbf{p}_1'-\frac{M_1\textbf{P}}{M_{\rm tot}}
=\frac{\textbf{p}_1'M_2-\textbf{p}_1'M_1}{M_{\rm tot}} \ .
\label{eq_CMback}
\end{align}
In vacuum, solving the equation in the CM frame and transforming back to an arbitrary 
frame results in the same solution as obtained from solving the original equation, due 
to Galilean invariance. No approximations are necessary in this procedure. 
In medium, neglecting the blocking factor and using the two-body selfenergy~\cite{Liu:2017qah} to include medium 
effects, the  $T$-matrix equation in the the-body CM frame is given by
\begin{align}
&T(E_\text{cm},\textbf{p}_\text{cm},\textbf{p}_\text{cm}')=
\nonumber\\
&V(\textbf{p}_\text{cm}-\textbf{p}_\text{cm}')+\int_{-\infty}^{\infty}
\frac{d^3\textbf{k}_\text{cm}}{(2\pi)^3}V(\textbf{p}_\text{cm}-\textbf{k}_\text{cm})
\nonumber\\
&\times\frac{1}{E_\text{cm}-\frac{(\textbf{k}_\text{cm})^2}{2\mu}
-\Sigma^{(2)}(E,\textbf{P},\textbf{k}_\text{cm})}
T(E_\text{cm},\textbf{k}_\text{cm},\textbf{p}_\text{cm}') \ .
\label{eq_NRTmCMwithSelfE}
\end{align}
Here, the CM approximation assumes that the two-body selfenergy only depends on 
\(\textbf{P}\) and \(E\) through \(E_\text{cm}\), so that 
\(\Sigma^{(2)}(E,\textbf{P},\textbf{k}_\text{cm})\approx
\Sigma^{(2)}(E_\text{cm},0,\textbf{k}_\text{cm})\equiv
\Sigma^{(2)}(E_\text{cm},\textbf{k}_\text{cm})\). The CM transformations  have the same form in
medium and in vacuum, but it is an approximation for the in-medium case. Thus, the CM 
transformation can be understood as expressing \(\textbf{p}_\text{cm}\) as a function of 
\(\{M_1,M_2,\textbf{p}_1,\textbf{p}_2\}\) and \(\textbf{p}_\text{cm}'\) as a function of 
\(\{M_1',M_2',\textbf{p}_1',\textbf{p}_2'\}\). This is the motivation for defining and 
analogous transformation for the relativistic in-medium off-shell case.

In the relativistic case, transformations to an arbitrary frame are Lorentz transformations 
(with $ \parallel $ and $ \perp $ indicating parallel and perpendicular to 
the relative velocity, respectively),
\begin{align}
&\varepsilon_p'=\gamma(\varepsilon_p-{v} {p}_\parallel) \ , \ 
p_{\parallel}'=\gamma(p_\parallel-v \varepsilon_p) \ , \ 
\textbf{p}_\perp'=\textbf{p}_\perp
\nonumber\\
&p_\parallel=\textbf{p}\cdot\hat{\textbf{v}} \ ,\ \textbf{p}_\perp=\textbf{p}-p_\parallel\hat{\textbf{v}},
\label{eq_lorentzoriginal}
\end{align}
where $\hat{\textbf{v}}$ denotes the unit vector in the direction of the velocity.  
Relativistic CM transformations, in analogy to Eq.~(\ref{eq_CMback}), are realized 
using the quantities
\begin{align}
&\textbf{v}_\text{cm}=\frac{\textbf{p}_1+\textbf{p}_2}{\varepsilon_{\text{p}_1}+\varepsilon_{\text{p}_2}}
\\
&\gamma_{\text{cm}}=\frac{\varepsilon_{\text{p}_1}+\varepsilon_{\text{p}_2}}{\sqrt{s}}
\\
&s=(\varepsilon_{\text{p}_1}+\varepsilon_{\text{p}_2})^2-(\textbf{p}_1+\textbf{p}_2)^2 \ .
\label{eq_vandgamma}
\end{align}
After obtaining the $T$-matrix solution in the CM frame, it is necessary to express 
$\{E_\text{cm}, \textbf{p}_\text{cm}, \textbf{p}'_\text{cm}\}$ in terms of 
$\{\textbf{p}_1,\textbf{p}_2, \textbf{p}_1',\textbf{p}_2',E\}$ to obtain the solution 
in an arbitrary frame. 
The relativistic CM transformation for energy is simply 
$\sqrt{s}\equiv E_\text{cm}=\sqrt{E^2-P^2}$. For the a CM 3-momentum, 
\(\textbf{p}_\text{cm}\), it can be expressed in terms of components parallel and 
perpendicular to $ \textbf{v}_{\text{cm}} $ as
\begin{align}
&p_{\text{cm}\parallel}=\gamma_{\text{cm}}(p_\parallel-v_\text{cm} \varepsilon_p)
=\frac{\varepsilon_{\text{p}_2}p_{1\parallel}-\varepsilon_{\text{p}_1}p_{2\parallel}}{\sqrt{s}}
\nonumber\\
&\textbf{p}_{\text{cm}\perp}=\textbf{p}_\perp=\textbf{p}-p_\parallel\hat{\textbf{v}}_\text{cm}
=\frac{\textbf{p}_1 p_{2\parallel}-\textbf{p}_2 p_{1\parallel}}{|\textbf{p}_1+\textbf{p}_2|} \ , 
\label{eq_lorentzprac}
\end{align}
and likewise for primed momenta, together with the constraint on total momentum conservation,
 $ \textbf{p}_1 +\textbf{p}_2=\textbf{p}_1' + \textbf{p}_2' $. 
The CM transformation for $ p_\text{cm}=|\textbf{p}_\text{cm}| $, $ p_\text{cm}'=|\textbf{p}_\text{cm}'|$, and 
$ \cos(\theta_{\text{cm}})=\textbf{p}_\text{cm}\cdot \textbf{p}_\text{cm}'/(p_\text{cm}p_\text{cm}') $  
in Eq.~(\ref{eq_ampsq}) can be obtained using Eq.~(\ref{eq_lorentzprac}) and its primed 
counterpart. The Galilean CM 
transformations are recovered in the nonrelativistic limit. In the on-shell limit, the 
relativistic CM transformation used in Ref.~\cite{Riek:2010py} is recovered. However, since
the transformation introduced here does not involve the external energies, the analytical properties
of the transformed $T$-matrix are preserved more accurately, while in the prescription of 
Ref.~\cite{Riek:2010py} the Lorentz invariance of the Mandelstam variables is preserved. Since our 
focus here is on the low-momentum properties of the heavy quarks, in connection with off-shell effects, 
we choose as our default is the new prescription of Eq.~(\ref{eq_lorentzprac}).
In practice, the imaginary parts of the parton selfenergies calculated within the present prescription 
tend to be 10\% larger at their peak values compared to the previous prescription used in 
Ref.~\cite{Riek:2010py}.

\bibliography{refcnew}

\end{document}